%% file: main_ieee.tex
\documentclass[conference]{IEEEtran}
\pdfoutput=1    % For arXiv issues

\usepackage{xcolor}
\definecolor{wong-black}        {HTML}{000000}
\definecolor{wong-lightorange}  {HTML}{E69F00}
\definecolor{wong-lightblue}    {HTML}{56B4E9}
\definecolor{wong-green}        {HTML}{009E73}
\definecolor{wong-yellow}       {HTML}{F0E442}
\definecolor{wong-darkblue}     {HTML}{0072B2}
\definecolor{wong-darkorange}   {HTML}{D55E00}
\definecolor{wong-pink}         {HTML}{CC79A7}

\usepackage[accsupp]{axessibility}  % Improves PDF readability for those with disabilities.

\usepackage{url}
\usepackage{hyperref} % Working hyperlink (https://www.overleaf.com/learn/latex/Hyperlinks)

\hypersetup{
    colorlinks=true,
    citecolor=wong-green,
    linkcolor=wong-darkblue,
    filecolor=wong-pink,
    urlcolor=wong-black,
    pdfpagemode=FullScreen,
    }

\usepackage{cite}
\usepackage{amsmath,amssymb,amsfonts}
\usepackage{algorithmic}
\usepackage{graphicx}
\usepackage{booktabs}
\usepackage{textcomp}
\usepackage{caption}
\usepackage{subcaption}
\usepackage{tikz}
\usepackage{comment}
\usepackage[colorinlistoftodos]{todonotes}
\usepackage{multirow}
\usepackage[nolist, nohyperlinks, printonlyused]{acronym} % For consistent acronyms

\def\BibTeX{{\rm B\kern-.05em{\sc i\kern-.025em b}\kern-.08em
    T\kern-.1667em\lower.7ex\hbox{E}\kern-.125emX}}

\newcommand\nnfootnote[1]{  % Footnote without hyperref association (https://tex.stackexchange.com/questions/415625/avoiding-hyperref-warning-ignoring-empty-anchor)
  \begin{NoHyper}
  \renewcommand\thefootnote{}\footnote{#1}%
  \addtocounter{footnote}{-1}%
  \end{NoHyper}
}

\begin{document}

% -------------------------- TITLE -------------------------------

\title{Specialized text classification: an approach to classifying Open Banking transactions}

% -------------------------- AUTHORS -------------------------------

\author{\IEEEauthorblockN{Duc Tuyen Ta,
        Wajdi Ben Saad,
        Ji Young Oh}

    \IEEEauthorblockA{Data Science Team - Oney Bank - France}
     %   $\{$dtta, obensaad, jyoh$\}$@oney.com
    % \IEEEauthorblockA{\IEEEauthorrefmark{3}Open Banking Data Science Team, Oney, France\\}
}

\maketitle

\nnfootnote{\textasteriskcentered~All authors contributed equally}

% -------------------------- ACRONYMS -------------------------------

\begin{acronym}
    \acro{ml}[ML]{Machine Learning}
    \acro{cnn}[CNN]{Convolutional Neural Network}
    \acro{dl}[DL]{Deep Learning}
    \acro{ad}[OB]{Open Banking}
\end{acronym}

% -------------------------- ABSTRACT -------------------------------

\begin{abstract}
    \input{sections/0_abstract}

\end{abstract}

% -------------------------- KEYWORDS -------------------------------


% -------------------------- CONTENT -------------------------------

\input{sections/1_introduction}
\input{sections/2_related_work}

\input{sections/3_system_description}
\input{sections/4_evaluation}
\input{sections/5_conclusion}
% \input{sections/6_acknowledgment}

% -------------------------- REFERENCES -------------------------------

{\small
    \bibliographystyle{IEEEtran}
    \bibliography{references}
}

\end{document}

%% file: sections/0_abstract.tex
%Single paragraph up to 250 words. Mini-version of the paper that includes context, state of the art, why it is not good enough, the research question, the methods, the evaluation and conclusions.
With the introduction of the PSD2 regulation in the EU which established the Open Banking framework, a new window of opportunities has opened for banks and fintechs to explore and enrich Bank transaction descriptions with the aim of building a better understanding  of customer behavior, while using this understanding to prevent fraud, reduce risks and offer more competitive and tailored services.

And although the usage of natural language processing models and techniques has seen an incredible progress in various applications and domains over the past few years, custom applications based on domain-specific text corpus remain unaddressed especially in the banking sector.

In this paper, we introduce a language-based Open Banking transaction classification system with a focus on the french market and french language text.
The system encompasses data collection, labeling, preprocessing, modeling, and evaluation stages. Unlike previous studies that focus on general classification approaches, this system is specifically tailored to address the challenges posed by training a language model with a specialized text corpus (Banking data in the French context).
By incorporating language-specific techniques and domain knowledge, the proposed system demonstrates enhanced performance and efficiency compared to generic approaches.

%% file: sections/1_introduction.tex
\section{Introduction}
\label{sec:introduction}

The European Union's strategy for digitalizing the financial sector is built on four major pillars: (1) extensive reporting requirements, (2) strict data protection rules, (3) Open Banking to enhance competition, and (4) a legislative framework for digital identification. In this regard, Open Banking (OB), which is under the Second Payment Services Directive (PSD2~\footnote{Directive (EU) 2015/2366 of the European Parliament and of the Council of 25 November 2015 on payment services in the internal market, amending Directives 2002/65/EC, 2009/110/EC and 2013/36/EU; Regulation (EU) No 1093/2010; Repealing Directive 2007/64/EC, OJ of 23.12.2015, L 337/35. This directive has enabled the introduction of new banking products and services, promoting competition without compromising security.}), facilitates the provision of customers' banking data to third-party FinTech companies. This pillar plays a significant role in enhancing competition in the financial sector. With the access to customers's banking data, third party providers could create new financial products and services, leads to increased innovation and choice for customers.

In the context of Open Banking, bank transaction classification is an important task in organizing and understanding financial data. Along with the increase in the adoption of OB initiatives and the availability of vast amounts of transaction data, there is a growing need for accurate and efficient classification systems. In this paper, we present the development of a language-focused bank transaction classification system applied to Open Banking data from French customers.

The goal of this work is to describe a comprehensive bank transaction classification system, covering various stages from data collection and labeling to preprocessing, modeling, and evaluation. Unlike previous studies that have mainly focused on general classification approaches, our system is specifically tailored to the unique characteristics of Open Banking data in the French context. By addressing the specific challenges of Open Banking data and focusing on the language aspect, we aim to enhance the effectiveness of bank transaction classification systems.

% We address the language nuances and challenges associated with this domain, aiming to achieve improved classification performance.

% The contribution of our paper lies in providing a detailed description of the entire classification system, highlighting the steps involved and the methodologies employed at each stage. By addressing the specific challenges of Open Banking data and focusing on the language aspect, we aim to enhance the effectiveness of banking transaction classification systems.

% To establish the relevance and novelty of our work, we review existing literature on transaction classification, emphasizing the gaps that our system aims to fill. We discuss the limitations of previous approaches and highlight the need for a language-focused system designed explicitly for Open Banking data.

% By presenting our language-focused banking transaction classification system, we aim to contribute to the advancement of classification techniques in the context of Open Banking. We anticipate that our system's insights and methodologies will enable more accurate and efficient categorization of transaction data, thereby facilitating improved financial data analysis and decision-making processes.

The paper is organized into five sections. Section~\ref{sec:related_work} provides an overview of existing research in transaction classification, emphasizing the specific aspects that differentiate our work. Section~\ref{sec:sys_desc} presents a comprehensive outline of our classification system. Section~\ref{sec:experiments} presents the results of our system's performance on a banking transaction dataset.
%, comparing it with existing approaches and highlighting its advantages. 
Finally, in section~\ref{sec:conclusion}, a summary of our findings, the implications of our work, and potential future research are presented.

%% file: sections/2_related_work.tex
\section{Related Work}
\label{sec:related_work}
Recent advances in text classification have been achieved through the utilization of advanced language models trained on large corpora. However, despite these improvements, short text classification like bank transaction description, continues to present numerous challenges. This section provides an extensive view of the related works to effectively address these challenges.
% The existing work are dicussed~\ref{sec:SOTA_short}. Section~\ref{sec:SOTA_bank} then examines the application of these models to the banking transaction classification. Finally, the findings from the related work are summarized in Section~\ref{sec:SOTA_summary}, which offers insights into the state of research in this area and identifies potential direction for our study.

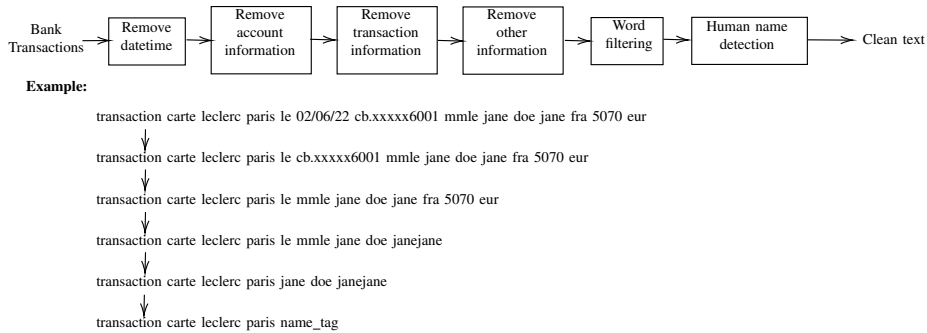
\begin{figure*}[t]
    \centering
    \resizebox{0.699\textwidth}{!}{
        \input{figs/preprocessing.tikz}
    }
    \caption{Illustration of the text preprocessing and human name detection process for banking transaction description.}
    \label{fig:system_preprocessing}
\end{figure*}

\subsection{Transaction Categorization}
\label{sec:SOTA_BT_classification}

Transaction classification involves identifying the purpose or context of a transactions using information such as description, amount, date, and metadata. Due to restricted access to these data, few research have concentrated on this problem~\cite{kotios2022deep,khazane2019deeptrax, vollset2017automatic, dayioglugil2017continuous, garcia2020identifying} where directly related are~\cite{kotios2022deep, vollset2017automatic, garcia2020identifying}.

In~\cite{kotios2022deep}, authors proposed general systems for bank transaction classification using machine learning models. A hybrid text classification method that combined rule-based and AI-based approaches is introduced. In~\cite{vollset2017automatic}, author explored the classification of bank transaction documents at Sparebank1 (Norway) using traditional classification methods like logistic regression and SVM. They employed a well-designed preprocessing process to remove sensitive information and non-alphabetic symbols. Additionally, they demonstrated that incorporating external resources, such as open data from government registries or the Google Places API~\footnote{\url{https://developers.google.com/maps/documentation/places/web-service/overview}}, could improve or reduce the accuracy of the classification model. In addition, the work in~\cite{garcia2020identifying} presented a classification method that employed preprocessing techniques, a specialized lexicon corpus, and multiple pairwise Support Vector Machines (SVM) classifiers. Their approach involved cleaning the text, anonymizing sensitive data, and constructing a labeled corpus with useful final elements. Input features such as lexical data, amount ranges, day ranges, word n-grams, and character n-grams were used for the SVM classifier. This work focused on Portuguese bank transactions.

\subsection{Summary}

\label{sec:SOTA_summary}
Bank transactions classification has received limited attention in the academic literature, primarily due to the sensitive nature of the data involved. Existing studies have primarily focused on data enrichment techniques, incorporating transaction-related information such as time and value or external information to improve classification accuracy. These studies often lack language specificity and require careful design of preprocessing methods to ensure adaptability to different languages.
% Assumptions are often made about the limited data compared to the number of transaction categories.
Furthermore, the approach of utilizing multiple pairwise classifiers becomes impractical when considering more detailed transaction type, such as distinguishing between food, fast-food, or restaurant transactions, due to the substantial number of possible categories.

Building upon the existing work, our study aims to contribute by exploring innovative techniques that leverage machine learning and advanced preprocessing approaches. Our goal is to enhance the performance of classification systems while simultaneously simplifying the overall system.
%Specifically, we will focus on analyzing banking transactions from French customers aggregated through the Open Banking platform, providing insights and advancements in this domain.

%% file: figs/preprocessing.tikz
\tikzset{every picture/.style={line width=0.75pt}} %set default line width to 0.75pt        

\begin{tikzpicture}[x=0.75pt,y=0.75pt,yscale=-1,xscale=1]
%uncomment if require: \path (0,317); %set diagram left start at 0, and has height of 317

%Straight Lines [id:da14517314215060084] 
\draw    (75,31) -- (97.4,30.86) ;
\draw [shift={(99.4,30.85)}, rotate = 179.65] [color={rgb, 255:red, 0; green, 0; blue, 0 }  ][line width=0.75]    (10.93,-3.29) .. controls (6.95,-1.4) and (3.31,-0.3) .. (0,0) .. controls (3.31,0.3) and (6.95,1.4) .. (10.93,3.29)   ;
%Shape: Rectangle [id:dp09010484242569738] 
\draw   (190.03,0.49) -- (279.43,0.49) -- (279.43,60.64) -- (190.03,60.64) -- cycle ;
%Straight Lines [id:da6854402133207886] 
\draw    (168,30.23) -- (188.4,30.79) ;
\draw [shift={(190.4,30.85)}, rotate = 181.6] [color={rgb, 255:red, 0; green, 0; blue, 0 }  ][line width=0.75]    (10.93,-3.29) .. controls (6.95,-1.4) and (3.31,-0.3) .. (0,0) .. controls (3.31,0.3) and (6.95,1.4) .. (10.93,3.29)   ;
%Straight Lines [id:da7542095633411263] 
\draw    (723,30.23) -- (761.4,30.51) ;
\draw [shift={(763.4,30.52)}, rotate = 180.43] [color={rgb, 255:red, 0; green, 0; blue, 0 }  ][line width=0.75]    (10.93,-3.29) .. controls (6.95,-1.4) and (3.31,-0.3) .. (0,0) .. controls (3.31,0.3) and (6.95,1.4) .. (10.93,3.29)   ;
%Straight Lines [id:da09023186404290606] 
\draw    (131.33,110.13) -- (130.98,126.93) ;
\draw [shift={(130.93,128.93)}, rotate = 271.22] [color={rgb, 255:red, 0; green, 0; blue, 0 }  ][line width=0.75]    (10.93,-3.29) .. controls (6.95,-1.4) and (3.31,-0.3) .. (0,0) .. controls (3.31,0.3) and (6.95,1.4) .. (10.93,3.29)   ;
%Straight Lines [id:da2082451957818212] 
\draw    (131.33,146.13) -- (130.98,162.93) ;
\draw [shift={(130.93,164.93)}, rotate = 271.22] [color={rgb, 255:red, 0; green, 0; blue, 0 }  ][line width=0.75]    (10.93,-3.29) .. controls (6.95,-1.4) and (3.31,-0.3) .. (0,0) .. controls (3.31,0.3) and (6.95,1.4) .. (10.93,3.29)   ;
%Straight Lines [id:da28236048007149583] 
\draw    (131.33,184.13) -- (130.98,200.93) ;
\draw [shift={(130.93,202.93)}, rotate = 271.22] [color={rgb, 255:red, 0; green, 0; blue, 0 }  ][line width=0.75]    (10.93,-3.29) .. controls (6.95,-1.4) and (3.31,-0.3) .. (0,0) .. controls (3.31,0.3) and (6.95,1.4) .. (10.93,3.29)   ;
%Straight Lines [id:da43592712845285675] 
\draw    (131.33,220.13) -- (130.98,236.93) ;
\draw [shift={(130.93,238.93)}, rotate = 271.22] [color={rgb, 255:red, 0; green, 0; blue, 0 }  ][line width=0.75]    (10.93,-3.29) .. controls (6.95,-1.4) and (3.31,-0.3) .. (0,0) .. controls (3.31,0.3) and (6.95,1.4) .. (10.93,3.29)   ;
%Straight Lines [id:da8768190316317499] 
\draw    (130.33,258.13) -- (129.98,274.93) ;
\draw [shift={(129.93,276.93)}, rotate = 271.22] [color={rgb, 255:red, 0; green, 0; blue, 0 }  ][line width=0.75]    (10.93,-3.29) .. controls (6.95,-1.4) and (3.31,-0.3) .. (0,0) .. controls (3.31,0.3) and (6.95,1.4) .. (10.93,3.29)   ;
%Shape: Rectangle [id:dp05898690874437462] 
\draw   (98.5,10.63) -- (166.9,10.63) -- (166.9,52.73) -- (98.5,52.73) -- cycle ;
%Shape: Rectangle [id:dp3174791998916344] 
\draw   (529.43,10.32) -- (593.7,10.32) -- (593.7,52.42) -- (529.43,52.42) -- cycle ;
%Shape: Rectangle [id:dp14937589042055355] 
\draw   (619.33,10.12) -- (722.6,10.12) -- (722.6,52.22) -- (619.33,52.22) -- cycle ;
%Shape: Rectangle [id:dp5808746333236203] 
\draw   (302.53,0.57) -- (391.93,0.57) -- (391.93,60.72) -- (302.53,60.72) -- cycle ;
%Straight Lines [id:da4845840961127814] 
\draw    (278.5,29.73) -- (298.9,30.29) ;
\draw [shift={(300.9,30.35)}, rotate = 181.6] [color={rgb, 255:red, 0; green, 0; blue, 0 }  ][line width=0.75]    (10.93,-3.29) .. controls (6.95,-1.4) and (3.31,-0.3) .. (0,0) .. controls (3.31,0.3) and (6.95,1.4) .. (10.93,3.29)   ;
%Shape: Rectangle [id:dp13721644941745326] 
\draw   (415.03,0.82) -- (504.43,0.82) -- (504.43,60.97) -- (415.03,60.97) -- cycle ;
%Straight Lines [id:da5103049453318989] 
\draw    (392.5,30.73) -- (412.9,31.29) ;
\draw [shift={(414.9,31.35)}, rotate = 181.6] [color={rgb, 255:red, 0; green, 0; blue, 0 }  ][line width=0.75]    (10.93,-3.29) .. controls (6.95,-1.4) and (3.31,-0.3) .. (0,0) .. controls (3.31,0.3) and (6.95,1.4) .. (10.93,3.29)   ;
%Straight Lines [id:da26310750184887755] 
\draw    (506,30.73) -- (526.4,31.29) ;
\draw [shift={(528.4,31.35)}, rotate = 181.6] [color={rgb, 255:red, 0; green, 0; blue, 0 }  ][line width=0.75]    (10.93,-3.29) .. controls (6.95,-1.4) and (3.31,-0.3) .. (0,0) .. controls (3.31,0.3) and (6.95,1.4) .. (10.93,3.29)   ;
%Straight Lines [id:da4921734706621681] 
\draw    (594,29.73) -- (614.4,30.29) ;
\draw [shift={(616.4,30.35)}, rotate = 181.6] [color={rgb, 255:red, 0; green, 0; blue, 0 }  ][line width=0.75]    (10.93,-3.29) .. controls (6.95,-1.4) and (3.31,-0.3) .. (0,0) .. controls (3.31,0.3) and (6.95,1.4) .. (10.93,3.29)   ;

% Text Node
\draw (0.67,14.15) node [anchor=north west][inner sep=0.75pt]   [align=left] {\begin{minipage}[lt]{60.73pt}\setlength\topsep{0pt}
\begin{center}
Bank\\Transactions
\end{center}

\end{minipage}};
% Text Node
\draw (102.97,12.68) node [anchor=north west][inner sep=0.75pt]   [align=left] {\begin{minipage}[lt]{41.85pt}\setlength\topsep{0pt}
\begin{center}
Remove\\datetime
\end{center}

\end{minipage}};
% Text Node
\draw (196.73,1.07) node [anchor=north west][inner sep=0.75pt]   [align=left] {\begin{minipage}[lt]{53.18pt}\setlength\topsep{0pt}
\begin{center}
Remove\\account\\information
\end{center}

\end{minipage}};
% Text Node
\draw (309.23,1.15) node [anchor=north west][inner sep=0.75pt]   [align=left] {\begin{minipage}[lt]{53.18pt}\setlength\topsep{0pt}
\begin{center}
Remove\\transaction\\information
\end{center}

\end{minipage}};
% Text Node
\draw (421.73,1.4) node [anchor=north west][inner sep=0.75pt]   [align=left] {\begin{minipage}[lt]{53.18pt}\setlength\topsep{0pt}
\begin{center}
Remove\\other\\information
\end{center}

\end{minipage}};
% Text Node
\draw (538.07,10.87) node [anchor=north west][inner sep=0.75pt]   [align=left] {\begin{minipage}[lt]{35.61pt}\setlength\topsep{0pt}
\begin{center}
Word\\filtering
\end{center}

\end{minipage}};
% Text Node
\draw (624.63,12.17) node [anchor=north west][inner sep=0.75pt]   [align=left] {\begin{minipage}[lt]{63.95pt}\setlength\topsep{0pt}
\begin{center}
Human name\\detection
\end{center}

\end{minipage}};
% Text Node
\draw (766,23.5) node [anchor=north west][inner sep=0.75pt]   [align=left] {\begin{minipage}[lt]{48.65pt}\setlength\topsep{0pt}
\begin{center}
Clean text
\end{center}

\end{minipage}};
% Text Node
\draw (18.67,65.67) node [anchor=north west][inner sep=0.75pt]   [align=left] {\begin{minipage}[lt]{48.08pt}\setlength\topsep{0pt}
\begin{center}
\textbf{Example:}
\end{center}

\end{minipage}};
% Text Node
\draw (86,92.13) node [anchor=north west][inner sep=0.75pt]   [align=left] {transaction carte leclerc paris le 02/06/22 cb.xxxxx6001 mmle jane doe jane fra 5070 eur};
% Text Node
\draw (86,129.06) node [anchor=north west][inner sep=0.75pt]   [align=left] {transaction carte leclerc paris le cb.xxxxx6001 mmle jane doe jane fra 5070 eur};
% Text Node
\draw (86,165.99) node [anchor=north west][inner sep=0.75pt]   [align=left] {transaction carte leclerc paris le mmle jane doe jane fra 5070 eur};
% Text Node
\draw (86,202.92) node [anchor=north west][inner sep=0.75pt]   [align=left] {transaction carte leclerc paris le mmle jane doe janejane};
% Text Node
\draw (86,239.85) node [anchor=north west][inner sep=0.75pt]   [align=left] {transaction carte leclerc paris jane doe janejane};
% Text Node
\draw (86,276.8) node [anchor=north west][inner sep=0.75pt]   [align=left] {transaction carte leclerc paris name\_tag};

\end{tikzpicture}

%% file: sections/3_system_description.tex
\section{System Description}
\label{sec:sys_desc}

To develop a simple and efficient bank transaction classification
system, our approach has three stages: (1) data preparation, (2) preprocessing and (3) modeling.

% \begin{figure}[t]
%     \input{figs/system_diagram.tikz}
%     \caption{Bank transaction classification stages.}
%     \label{fig:system_diagram}
% \end{figure}

\begin{table*}[t]
    \centering
    \caption{Example of sample transaction data entries with these corresponding categories. Limited by the data privacy restrictions, the data is masked by the corresponding text.}
    \label{table:example_trx}
    % \resizebox{\columnwidth}{!}{%
    \begin{tabular}{@{}llll@{}}
        \toprule
        Description                                                                                                                                                                                        & Value  & Date      & Category       \\
        \hline\hline
        030522 cb****1234 super\_market\_name city\_name                                                                                                                                                   & -29.55 & 3/5/2022  & GROCERIES      \\
        % 050422 cb****1234 hotel\_name rero cluj city\_name 290,00ron 1 euro = 4,885444                                                                                                                     & -59.36 & 03/04/2022 & TRAVEL           \\
        % 010621 cb****4554 clothing\_company city\_name                                                                                                                                                     & -95.8  & 16/12/2021 & CLOTHES \& SHOES \\
        \begin{tabular}[c]{@{}l@{}}virement en votre faveur avance salaires 10.2021 1000.0 carte numero 123\\ $\rightarrow$ \textit{ transfer advance salary 10.2021 by bank card number 123}\end{tabular} & 1000   & 9/11/2022 & ADVANCE SALARY \\
        091221 cb****1234 fastfood\_company country\_code city\_name 2,20eur 1 euro = 1,000000                                                                                                             & -2.2   & 9/12/2021 & FAST FOODS     \\
        \toprule
    \end{tabular}%
\end{table*}

\subsection{Data Preparing}
\label{sec:SYS_data}

The real customer's data is collected from the Open Banking platform with their consent. We have selected only relevant fields for the classification purpose. Section~\ref{sec:eval_ml} describes the resulting dataset which is used for modeling and evaluation.

\subsection{Preprocessing}
\label{sec:SYS_pre}
Bank transaction descriptions serve as concise representations of transactions, encompassing various elements such as the transaction date, customer information (e.g., human names), account details (e.g., card number, IBAN, etc.), and transaction-specific information (e.g., currency, rate). These descriptions condense the meaning of the transaction into a limited character length. Notably, these descriptions often lack verbs, emphasizing their brevity and condensed nature.

Despite their brevity, these descriptions often contain irrelevant or extraneous information and duplicated text which become a complex challenge in transaction classification. In this perspective, it is crucial to develop a preprocessing stage that is specifically designed to adapt to particular characteristics of Open Banking data in French context. This preprocessing stage should effectively handle the challenges associated with bank transaction descriptions, ensuring the removal of irrelevant information, identification of duplicate content, and mitigation of confusing text. By designing a preprocessing approach that addresses these specific challenges, we can improve the accuracy and reliability of subsequent classification tasks for Open Banking data in the French context.
% Moreover, duplicate information or confusing text can be present, introducing potential challenges in accurately categorizing transactions.
% For instance, giant French supermarket chains may offer additional financial support systems to customers during their product purchases. Although the transaction information may indicate a transaction with a supermarket, the actual transaction category could be completely different.

\subsubsection{Text cleaning}
\label{sec:SYS_pre_token}
To selectively preserve important information from meaningless text, the preprocessing stage with example is illustrated in Figure~\ref{fig:system_preprocessing}. It includes: 1, removing the date-time information in the transaction, e.g. "\textit{02/06/2022}"; 2, removing account-related information, e.g. the masked card number; 3, removing transaction information like currencies; 4, removing other information like the location, the gender, postal code. and 5, filtering stop words. Duplicates or words with same meanings such as "\textit{cb}", "\textit{carte bancaire}", "\textit{carte}" referring to all "\textit{bank cards}" are removed, retaining the first occurrence word.

% \begin{itemize}
%     \item Removing date time. To remove the date-time information in the transaction, e.g. "\textit{02/06/2022}", or "\textit{jun 2022}".
%     \item Removing account information. To remove the account-related information, e.g. the masked card number.
%     \item Removing transaction information. To remove the transaction information, like currencies, exchange rates, etc.
%     \item Removing other information. To remove other information like the location, the gender such as "\textit{mle}", etc.
%     \item Filtering stop words. To remove other useless information such as hash code, stop words, postal code, etc.
% \end{itemize}

\subsubsection{Human name anonymization}
\label{sec:SYS_pre_name}
Subsequently, the system identifies human names within the description by applying a name dictionary and replaces them with a designated tag. This particular step holds significance in constructing the word corpus, as transactions containing human names, such as bank transfers or card transactions, comprise more than $10\%$ of the total transactions.

% Again, the system effectively removes duplicated words from the transaction text, retaining only the first occurrence of each word.

\subsubsection{Similarity Filtering}
\label{sec:SYS_pre_sim}
Many transactions exhibit duplication in meaning, with slight variations in human names, locations, or additional details. It becomes imperative to incorporate a similarity detector in preparing the dataset for model training. However, with a scale of millions of samples, traditional methods using pairwise distance measurements, such as in~\cite{garcia2020identifying} with Jaccard distances, would be impractical. In this paper, we therefore adopt a different approach by transforming the transaction descriptions into vector form using Term Frequency-Inverse Document Frequency (TF-IDF) as terms to identify similar strings. This transformation allows us to reform the problem as a matrix multiplication task, which offers significantly improved computational efficiency. By leveraging TF-IDF, we can effectively compute the cosine similarity between these vectors, enabling the identification of similar data instances. Our experiments demonstrate that this process can be accomplished within one hour using our local computation environment, as outlined in Section~\ref{sec:experiments}. We then used this approach to prepare a dataset of more than $200,000$ data entries, divided into training and testing sets, from $5$ million transactions randomly sampled from the database.
% To perform the sparse matrix multiplication, we used $\textit{sparse\_dot\_topn}$~\footnote{\url{https://github.com/ing-bank/sparse_dot_topn}}, a package that provides a fast way to perform a sparse matrix multiplication followed by top-\textit{n} multiplication result selection.

% By leveraging TF-IDF, we can effectively compute the cosine similarity between these vectors, enabling the identification of similar data instances. This transformation allows us to reform the problem as a matrix multiplication task, which offers significantly improved computational efficiency. 

% The similarity filtering system is presented in Figure~\ref{fig:system_filtering} in which the rule-based classifier is explained in~\ref{sec:SYS_model_rule}.
% \begin{figure*}[t]
%     \resizebox{0.9\textwidth}{!}{
%         \input{figs/similarity.tikz}
%     }
%     \caption{Similarity filtering method to create the train and test dataset.}
%     \label{fig:system_filtering}
% \end{figure*}

\subsection{Modeling}
\label{sec:SYS_model}

\subsubsection{Feature Selections}
\label{sec:SYS_ model_token}
% Based on the structure of transaction entries, the possible features which can be considered in our classification system are the following:
% \begin{enumerate}
%     \item \textbf{Word n-grams}. N-gram representation of the transaction at the word level. %We applied the word n-gram transformation for the preprocessed text.
%     \item \textbf{Character n-grams}. N-gram representation of the transaction at the character level. %Again we applied the character n-gram transformation for the preprocessed text.
%     \item \textbf{Value}. The value of the transaction
%     \item \textbf{Date}. The date of the transaction.
% \end{enumerate}

Within the framework of this article, we focus on using only text descriptions to classify transactions. Specifically, word \textit{n}-grams are used as a feature to vectorize the preprocessed transaction text.

\subsubsection{Rule-based Labelling}
\label{sec:SYS_model_rule}
Instead of relying on manual labeling of individual transactions, our approach employs a rule-based method to label the dataset. This technique uses specific keywords present within the transaction information and considers the transaction value (income or expense) to assign labels to transactions. Each category is defined by a set of inclusive and exclusive keywords. When a transaction contains the inclusive keywords and does not include the excluded keywords associated with a particular category, the corresponding label is assigned. To ensure accuracy, this method is applied to the original transaction descriptions, rather than the preprocessed descriptions. The resulting labels are manually validated, enabling the creation of both training and testing datasets for modeling purposes.

\subsubsection{ML-based Classifier}
\label{sec:SYS_model_ml}

Considering the extensive range of categories involved in the bank transaction classification system, employing multiple one-versus-one pairwise classifiers, as demonstrated in~\cite{garcia2020identifying}, becomes increasingly impractical. This approach requires the creation of individual classifiers for each pair of categories, posing challenges in terms of scalability and computational resources.

To overcome these limitations, using a multi-class classifier is a more practical alternative. This approach allows for the simultaneous processing of numerous categories in a streamlined and efficient manner. By employing a single classifier to classify transactions across multiple categories, we simplify the classification process, eliminating the need for intricate pairwise comparisons, and reducing the computational burden.

%% file: sections/4_evaluation.tex
\section{Experiments}
\label{sec:experiments}

All experiments were performed on a computer with Intel@Core i7 11850 CPU 2.5GHz $\times$ 8, 15.4 Gb RAM and 476 Gb Disk, and Windows 10 Enterprise.

% \begin{table}[t]
%   \centering
%   \caption{Distribution of sample categories in the training dataset.}
%   \label{table:distribution}
%   \begin{tabular}{@{}l|l@{}}
%     Category            & Instances \\
%     \hline
%     GROCERIES           & 11,532    \\
%     WITHDRAWALS         & 8,238     \\
%     OTHER TRANSFER      & 6,772     \\
%     % GENERAL INSURANCE   & 6,133     \\
%     SALARY              & 5,649     \\
%     TELECOM \& INTERNET & 4,586     \\
%     LOAN                & 3,945     \\
%     MONEY TRANSFER      & 3,875     \\
%     INCIDENT FEES       & 2,354     \\
%     FAST FOODS          & 1,798     \\
%     TRAVEL              & 178       \\
%     CREDIT FEES         & 65        \\
%     ADVANCE SALARY      & 6         \\
%   \end{tabular}
% \end{table}

\begin{table}[t]
  \caption{Experiment results of various training data sizes and machine learning models.}
  \label{tab:exp_1}
  \begin{tabular}{lllll}
    Train                 & Model                 & \textbf{Precision} & \textbf{Recall} & \textbf{F$_1$} \\
    \hline\hline
    \multirow{5}{*}{80\%} & Logistic Regression   & 92\%               & 92\%            & 92\%           \\
                          & Random Forest         & 89\%               & 88\%            & 88\%           \\
                          & Linear SVM            & 94\%               & 94\%            & 94\%           \\
                          & Naives Bayes          & 88\%               & 86\%            & 85\%           \\
                          & Fine-tuned Linear SVM & \textbf{94\%}      & \textbf{94\%}   & \textbf{94\%}  \\
    \hline
    \multirow{5}{*}{67\%} & Logistic Regression   & 90\%               & 89\%            & 88\%           \\
                          & Random Forest         & 86\%               & 83\%            & 83\%           \\
                          & Linear SVM            & 92\%               & 91\%            & 91\%           \\
                          & Naives Bayes          & 86\%               & 83\%            & 82\%           \\
                          & Fine-tuned Linear SVM & \textbf{93\%}      & \textbf{92\%}   & \textbf{92\%}  \\
    \hline
    \multirow{5}{*}{50\%} & Logistic Regression   & 88\%               & 86\%            & 85\%           \\
                          & Random Forest         & 82\%               & 76\%            & 76\%           \\
                          & Linear SVM            & 90\%               & 89\%            & 89\%           \\
                          & Naives Bayes          & 84\%               & 80\%            & 79\%           \\
                          & Fine-tuned Linear SVM & \textbf{91\%}      & \textbf{90\%}   & \textbf{90\%}
  \end{tabular}%
\end{table}

\subsection{Dataset}
\label{sec:eval_data}
The data consists of a training set with $94,356$ transactions and a test set with $109,509$ transactions, both from customer accounts of major French banks, written mostly in French, from December $2021$ to January $2023$.

% . They were collected during the period from December $2021$ to January $2023$ through the Open Banking platform and sampled after the similarity filtering step as discussed in~\ref{sec:SYS_pre_sim}.

The entries of the dataset have the following attributes: i) \textbf{description}: the transaction description, ii) \textbf{value}: the transaction value in euros and iii) \textbf{date}: the date when the transaction occurred. Every entry has an extra category label that determines the classification goal. There are a total of $84$ categories, corresponding to different types of transactions in detail.
% The numerical distributions of several categories in the training dataset are presented in Table~\ref{table:distribution}. 
An example of data entries with their corresponding category is shown in Table~\ref{table:example_trx}. Due to the nature of the bank transaction, some categories appear frequently like ``GROCERIES'', periodically like ``SALARY'', or rarely like ``ADVANCE SALARY''. Consequently, the data set is heavily imbalanced, with some categories appearing only in a few.
% For example, "SALARY" is for regular salary payments while "ADVANCE SALARY" is for advance salary payments. The detailed level of categories helps to build a specific database about customer behavior, then to develop and building "custom" financial solution for each customer, as part of the Open Banking orientation.

% The numerical distributions of several categories in the training dataset are presented in Table~\ref{table:distribution}. An example of data entries with their corresponding category is shown in Table~\ref{table:example_trx}. Due to the nature of the banking transaction, some categories appear frequently like ``GROCERIES'', periodically like ``SALARY'', or rarely like ``ADVANCE SALARY''. Consequently, the data set is heavily imbalanced, with some categories appearing only in a few.

\subsection{Metrics}
\label{sec:eval_metrics}
Due to the heavy imbalance of the dataset as well as the large number of categories, we employed precision, recall, and $F_1$ metrics using a weighted-average approach. It means we will assign greater contributions to classes with more examples in the dataset, or the contribution of each class to the precision, recall, and $\text{F}_1$ average is weighted by its size.

\subsection{ML-based}
\label{sec:eval_ml}
To train the model, we consider different dataset splits, where the training data contains $80\%$, $67\%$, and $50\%$ of the original training set. The general performance of the model is then evaluated in the test dataset. The purpose is to test the robustness of the model when there is less training data available while also ensuring separation between the data used to train and to evaluate the model that is faced in real use cases.

\subsubsection{Experiment 1}

Each test uses the word \textit{n}-grams of the preprocessed transaction description as the input feature. For our experiment, $n=3$ is selected. The transformed word \textit{n}-grams then is vectorized by using the TF-IDF.

Table~\ref{tab:exp_1} shows the results with various training data sizes and machine learning models. We observe with only transaction description, the system with Linear SVM model can get the (weighted) precision, recall, and $F_1$ of $90\%$ with only $50\%$ of the training set and $94\%$ with $80\%$ of the training set.

\subsubsection{Experiment 2}

In this experiment, the preprocessed transaction description is vectorized using Word2Vec. The corpus is built using tokens from the training dataset. Sequence padding is applied to unify the length of each transaction text with a length of $14$. In the experiment, different sizes of embedding vectors are considered. To reduce the size of the training vector, Principal Component Analysis (PCA) is then applied. The number of selected components for PCA is $300$, ensuring that the output vector retains at least $98\%$ of the information present in the original vector.

Table~\ref{tab:exp_2} shows the results. We observe that with Word2Vec embedding approach, the training model with Random Forest can get the (weighted) precision, recall, and $F_1$ of $95\%$ in the tests set. Table~\ref{tab:exp_2_by_cat} shows the precision, recall, and $\text{F}_1$ score for several transaction categories by using Random Forest and Word2Vec embedding approach. We obtained overall high performance. The worst performance corresponds to the categories with fewer entries in the training set. In the case of "GROCERIES", the classification highly relied on the merchant name which explains the relatively low performance. To improve the performance for these categories, one of the future solutions is to use external data, such as a merchant database as in~\cite{vollset2017automatic}, to make decisions.

\begin{table}[t]
  \caption{Experiment results of various Word2Vec vector size and ML models.}
  \label{tab:exp_2}
  \begin{tabular}{llllll}
    \multicolumn{2}{l}{Word2Vec}                          & Model               & \textbf{Precision} & \textbf{Recall} & \textbf{F$_1$} \\
    \hline\hline
    \multicolumn{2}{l}{\multirow{3}{*}{vector\_size=300}} & Logistic Regression & 82\%               & 82\%            & 81\%           \\
    \multicolumn{2}{l}{}                                  & Random Forest       & \textbf{95\%}      & \textbf{94\%}   & \textbf{94\%}  \\
    \multicolumn{2}{l}{}                                  & Linear SVM          & 84\%               & 78\%            & 76\%           \\
    \hline
    \multicolumn{2}{l}{\multirow{3}{*}{vector\_size=200}} & Logistic Regression & 80\%               & 80\%            & 79\%           \\
    \multicolumn{2}{l}{}                                  & Random Forest       & \textbf{95\%}      & \textbf{94\%}   & \textbf{94\%}  \\
    \multicolumn{2}{l}{}                                  & Linear SVM          & 83\%               & 77\%            & 75\%           \\
    \hline
    \multicolumn{2}{l}{\multirow{3}{*}{vector\_size=100}} & Logistic Regression & 81\%               & 81\%            & 80\%           \\
    \multicolumn{2}{l}{}                                  & Random Forest       & \textbf{95\%}      & \textbf{94\%}   & \textbf{94\%}  \\
    \multicolumn{2}{l}{}                                  & Linear SVM          & 80\%               & 74\%            & 71\%
  \end{tabular}%
\end{table}

\begin{table}[t]
  % Please add the following required packages to your document preamble:
  % \usepackage{graphicx}
  \caption{Weighted precision, recall and $F_1$ score for several categories with Random Forest and Word2Vec.}
  \label{tab:exp_2_by_cat}
  \begin{tabular}{lllll}
                        & \multicolumn{1}{c}{\textbf{Precision}} & \multicolumn{1}{c}{\textbf{Recall}} & \multicolumn{1}{c}{\textbf{$\text{F}_1$}} & \multicolumn{1}{c}{\textbf{Support}} \\
    \hline\hline
    GROCERIES           & 87\%                                   & 97\%                                & 92\%                                      & 11726                                \\
    WITHDRAWALS         & 100\%                                  & 100\%                               & 100\%                                     & 8856                                 \\
    OTHER TRANSFER      & 90\%                                   & 97\%                                & 93\%                                      & 7692                                 \\
    % GENERAL INSURANCE   & 90\%                                   & 99\%                                & 94\%                                            & 7202                                 \\
    SALARY              & 91\%                                   & 97\%                                & 94\%                                      & 6098                                 \\
    TELECOM \& INTERNET & 99\%                                   & 97\%                                & 98\%                                      & 4550                                 \\
    LOAN                & 99\%                                   & 97\%                                & 98\%                                      & 6107                                 \\
    MONEY TRANSFER      & 99\%                                   & 99\%                                & 99\%                                      & 6038                                 \\
    INCIDENT FEES       & 100\%                                  & 100\%                               & 100\%                                     & 2438                                 \\
    FAST FOODS          & 98\%                                   & 91\%                                & 94\%                                      & 1809                                 \\
    TRAVEL              & 97\%                                   & 81\%                                & 89\%                                      & 272                                  \\
    CREDIT FEES         & 94\%                                   & 76\%                                & 84\%                                      & 84                                   \\
    ADVANCE SALARY      & 77\%                                   & 91\%                                & 83\%                                      & 11
  \end{tabular}%
\end{table}

Overall, these experimental results show the system's ability to accurately classify transactions, achieving high precision and recall rates across multiple transaction categories. This demonstrates the practicality and efficiency of the proposed system in real-world banking environments.

%% file: sections/5_conclusion.tex
\section{Conclusion}
\label{sec:conclusion}

This paper presented a comprehensive system for banking transaction classification with Open Banking data in France, which aims to accurately categorize various types of transactions. The proposed system demonstrated promising results, showing its effectiveness in achieving accurate classification outcomes. By leveraging a multi-class classifier, our system offers a simpler and more streamlined approach compared to existing methods. This approach eliminates the need for complex feature engineering or multiple stages of classification, while still achieving commendable performance. While the proposed system has demonstrated promising results, further research and validation are necessary to address challenges such as imbalanced data and evolving transaction types.

% The experimental results showcased the system's ability to classify transactions accurately, with high precision and recall rates across multiple transaction categories.

% These findings indicate the potential practicality and efficiency of the proposed system in real-world banking environments. The simplicity of the system also presents advantages in terms of ease of implementation and maintenance. It reduces computational complexity, making it more accessible to users with limited computational resources. Moreover, the system's straightforward design facilitates future enhancements and adaptability to evolving banking systems and transaction patterns.

% While the proposed system has demonstrated promising results, further research and validation are necessary to ensure its robustness and generalizability across diverse banking datasets and environments. Additionally, future work could focus on exploring techniques to address challenges such as imbalanced data, evolving transaction types, and potential security vulnerabilities.

% Overall, this paper contributes to the field of banking transaction classification by presenting a simplified yet effective system. The research findings highlight the potential for improved transaction categorization, offering practical benefits to financial institutions, regulators, and customers alike.